# Optimizing the use of pressurized bladders for the assembly of HL-LHC MQXFB magnets

J. Ferradas Troitino[a], G. Ambrosio[b], N. Bourcey[a], D. Cheng[c], A. Devred[a], H. Felice[d], P. Ferracin[c], M. Guinchard[a], S. Izquierdo Bermudez[a], K. Kandemir[a], N. Lusa[a], A. Milanese[a], S. Mugnier[a], J.C. Perez[a], E. Todesco[a], S. Triquet[a], G. Vallone[c]

[a] *CERN, the European Center for Nuclear Research. Geneva 23, 1211, Switzerland*
[b] *Fermi National Accelerator Laboratory, Batavia, IL 60510, USA*
[c] *Lawrence Berkeley National Laboratory, Berkeley, CA 94720, USA*
[d] *IRFU, CEA, Université ́ Paris-Saclay, F 91191 Gif-sur-Yvette, France*

**Abstract**

The use of pressurized bladders for stress control of superconducting magnets was firstly proposed at Lawrence Berkeley National Laboratory (LBNL) in the early 2000s. Since then, the so-called "bladders and keys" procedure has become one of the reference techniques for the assembly of high-field accelerator magnets and demonstrators. Exploiting the advantages of this method is today of critical importance for $Nb_3Sn$-based accelerator magnets, whose production requires the preservation of tight stress targets in the superconducting coils to limit the effects of the strain sensitivity and brittleness of the conductor.

The present manuscript reports on the results of an experimental campaign focused on the optimization of the "bladders and keys" assembly process in the MQXFB quadrupoles. These 7.2 m long magnets shall be among the first $Nb_3Sn$ cryomagnets to be installed in a particle accelerator as a part of the High Luminosity upgrade of the LHC. One of the main practical implications of the bladders technique, especially important when applied to long magnets like MQXFB, is that to insert the loading keys, the opening of a certain clearance in the support structure is required. The procedure used so far for MQXF magnets involved an overstress in the coils during bladder inflation. The work presented here shows that such an overshoot can be eliminated thanks to additional bladders properly positioned in the structure. This optimized method was validated in a short model magnet and in a full-length mechanical model, becoming the new baseline for the series production at CERN. Furthermore, the results are supported by numerical predictions using Finite Element models.

## 1. Introduction

### 1.1. Mechanical Design of Accelerator Magnets

During operation, the coil windings of superconducting magnets are subjected to strong electro-magnetic (e.m.) forces. The latter arise from the combination of large electrical transport currents and the high magnetic fields produced by the magnets.

With the objective of reacting these loads, the coils of accelerator magnets are assembled inside a mechanical structure, whose purpose is twofold. First, it is conceived to act as a force-restraining structure. Second, the structure has as well the essential role of minimizing the conductor displacement or deformation during magnet powering, ensuring the design field quality and reducing the potential risk of premature quenches caused by cracking or by frictional movements in the system.

How to create an optimum mechanical support for the superconducting coils (fulfilling the previous objectives) is a topic extensively treated in the magnet community, which ultimately leads to the so-called pre-stress or pre-load concept [1]-[4]. Both terms are equivalent and refer to the application of an initial and controlled load to the superconducting coils, aimed at counteracting the effect of the e.m. forces. As an example, in saddle-shape $\cos(\vartheta)$ dipole and $\cos(2\vartheta)$ quadrupole accelerator magnets, the Lorentz forces acting in the coils have three main components [4]: (1) an azimuthal component, tending to compress the coil towards its midplane, (2) a radial component, tending to push the coil outwardly, with a maximum effect towards the midplane, and (3) a longitudinal component, concentrated over the end coils, tending to elongate the coil. To compensate for the effects of these forces, a coil pre-load can be applied during the magnet assembly phase in the form of an initial pre-compression in the different spatial directions. Thanks to such counterbalancing forces, the deformations/displacements in the system during magnet powering can be reduced. Ideally, a full magnet pre-load would be such as to ensure that the pre-load force value equals the one of the total electro-magnetic force.

### 1.2. $Nb_3Sn$ Accelerator Magnets

The importance of the coil pre-stress increases as we move towards higher field magnets, with larger stored energies and e.m. loads. Up to the Large Hadron Collider magnets (LHC) at CERN, the preferred superconductor technology for dipole and quadrupole magnets was Nb–Ti. For the High Luminosity LHC (HL-LHC) upgrade project [5], the field and field gradient requirements impose to switch to Nb$_3$Sn (Niobium-Tin) technology for some of the magnets. Nb$_3$Sn conductors can operate at higher fields than Nb-Ti conductors and have become today the preferred technological choice for high-field accelerator magnets. As an example, for the Nb$_3$Sn RRP 108/127 HL-LHC wires, the minimal critical current requirement is 1280 A mm$^{-2}$ at 15 T and 4.22 K [6]. Thus, Nb$_3$Sn technology seeks to unlock the previous magnetic field frontiers and to allow the design of accelerator magnets operating above 10 T [7]. A critical aspect of Nb$_3$Sn superconductors is, however, their mechanical strain tolerance. The critical current density of Nb$_3$Sn cables depends on the strain applied to the superconducting phase [8]-[13], this last being, in addition, a brittle intermetallic compound with a noteworthy propensity to fracture. If the loads experienced by the conductor exceed a certain threshold, $J_c$ can be permanently reduced due to the fractures occuring in the Nb$_3$Sn phase and/or due to the presence of residual mechanical strain in the stabilizing matrix [14]-[20].

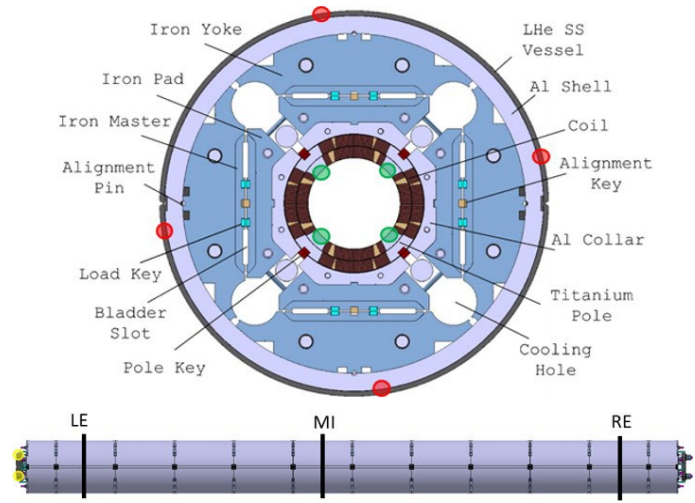

Figure 1: MQXFB magnet cross section and its side view. The position of the mechanical instrumentation is indicated by the round markers. The strain gauges placed in the Al-cylinder are removed after magnet assembly, prior to the preparation of the cold-mass. The original bladder slots are shown using their corresponding label. The external Al-cylinder is commonly referred to as "shell".

*1.3. Peak Stress Limits on Nb$_3$Sn Superconductors*

Respecting the mechanical limits at which Nb$_3$Sn superconductors can safely operate is hence crucial, and requires a thorough determination of the stress/strain values that the conductor can withstand under loads applied in different directions. Coil peak stress limits are in consequence set at both room (during assembly) and at cryogenic temperature (after cool-down and during energization). Their exact value depends on the particular cable employed and the load direction, for instance, the limits used for MQXFA&B programs (see next section) were initially in the 150 MPa to 200 MPa range [6], [21]-[24]. These thresholds apply for transverse stresses applied to superconducting cables (transverse to the cable broad face). A number of recent studies carried out at CERN for MQXF cables indicate that micro-cracking may initiate in the Nb$_3$Sn phases of wires subjected to transverse stresses at earlier levels, in the 100 MPa to 150 MPa range at room temperature [25]-[26]. Equally of concerns, are the tensile stresses that may develop under the axial component of the Lorentz load in the coils, but these are addressed by different features in the mechanical design and are not discussed in this paper.

*1.4. Quadrupole Magnets for HL-LHC Interaction Regions*

The first milestone for Nb$_3$Sn applications in High Energy Physics (HEP) shall be achieved with the installation of the Niobium-Tin quadrupole magnets, MQXF, built for the High-Luminosity upgrade of the LHC (HL-LHC) at CERN [5].

The mechanical assembly of these magnets is based on the so-called Bladder and Key (B&K) process pioneered at LBNL in the early 2000s [27]-[28], where the superconducting coils are assembled inside a support structure relying on aluminum shrinking cylinders (see Figure 1). The desired mechanical pre-load to the coils is applied through a system of water-pressurized bladders and interference keys. When water is pumped through the bladders, the provoked expansion results into the pre-tensioning of the Aluminum shrinking cylinder. The balance between the shrinking cylinder azimuthal tension and the coil azimuthal compression is then kept via the insertion of interference keys. The "bladdering and keying" operations are usually performed in steps, gradually increasing the key size until the desired coil pre-load at ambient temperature is obtained. Once the magnet loading is completed, an outer stainless-steel shell is welded around the magnet structure to form the helium containment vessel. The design and implementation of the stainless-steel shell has been iterated to limit the mechanical coupling between the outer shell and the magnet structure [29]. During cool-down, the pre-stress level is expected to further increase up to the target value at cryogenic temperature thanks to the selection of Aluminum for the shrinking cylinder, which presents a larger thermal contraction than the rest of the magnet components. The amount of pre-stress increase, from warm to cold, is related to the aperture of the magnet, the diameter of the shell and its thickness. The decoupled outer stainless steel shell is expected to not play any role.

To better illustrate the principle behind the shrinking cylinder-based support structure, the computed (using a 2D Finite Element model) strain, stress and force values for MQXF magnet after room temperature pre-load and at cryogenic temperature are summarized in Table 1. The first column provides the calculated peak azimuthal stress in the inner layer pole turn conductor, $\sigma_{\vartheta\ pole\ turn\ IL}$. The two next columns indicate the expected local stress values on the coil winding pole, $\sigma_{\vartheta\ pole}$, and on the Al-cylinder, $\sigma_{\vartheta\ cyl.}$, at the position of the mechanical instrumentation placed in the magnet (see Section 2). The following two columns provide the average stress values on the complete superconducting coil area (Avg. $\sigma_{\vartheta\ coil}$, all coil conductor blocks) and on the Al-cylinder surface (Avg. $\sigma_{\vartheta\ cyl.}$). The last three columns summarize the azimuthal force balance in the system ($F_{\vartheta\ pole}$, $F_{\vartheta\ cyl.}$) and the force ratio against the total electro-magnetic force ($F_{pole}/F_{e.m}$). As it can be seen, the Al-cylinder provides the azimuthal force pre-loading the coils. This force is doubled when the magnet is cooled-down to cryogenic temperature due to the differential thermal contraction between the external cylinder and the rest of the magnet components. The same effect is seen for the average coil and Al-cylinder stresses at 1.9 K: the average coil azimuthal pre-stress at cryogenic temperature is twice the one at room temperature. However, it is important to notice that when looking at the local winding pole stress values (at the position of the mechanical instrumentation), the increase of pre-load is in the order of 30 %. The explanation to this behavior can be found in the fact that the values reported from the model (or measured by the mechanical instrumentation) in the winding pole correspond to a local strain/stress field. The detailed magnet geometry, in combination with non-linear frictional and bending effects, create a non-uniform stress distribution across the mechanical structure and superconducting coils. As a result, the local winding pole value provides a good estimation of the peak azimuthal stress seen in the pole turn conductor of the inner layer. However, it does not give a complete insight into the average pre-load in the magnet coils.

Table 1: Computed stress and force values using the reference MQXF 2D magnet Finite Element model.

| Step | Max. $\sigma_{\vartheta\ pole\ turn\ IL}$ [MPa] | $\sigma_{\vartheta\ pole}$ [MPa] | $\sigma_{\vartheta\ cyl.}$ [MPa] | Avg. $\sigma_{\vartheta\ coil}$ [MPa] | Avg. $\sigma_{\vartheta\ cyl.}$ [MPa] | $F_{\vartheta\ pole}$ [kN m$^{-1}$] | $F_{\vartheta\ cyl.}$ [kN m$^{-1}$] | $F_{pole}/F_{e.m}$ % |
|---|---|---|---|---|---|---|---|---|
| After RT pre-load | -86 | -80 | 53 | -34 | 49 | 1530 | 1540 | 40 |
| At 1.9 K | -113 | -118 | 125 | -73 | 120 | 3310 | 3220 | 87 |

*1.5. Collaring vs. Bladder and Key*

The B&K assembly technique was developed as an alternative to the existing collaring strategy, which calls for heavy tooling such as a collaring press [30]-[31]. In collared coil structures, as it is the case of the Nb-Ti main quadrupole and dipole magnets for the LHC, structural collars (assembled in packs) are mounted and locked by means of keys around the superconducting coils. The pre-load is applied by a geometrical interference between the collar cavity and the coils. Here, the assembly is performed under a precisely aligned hydraulic press that allows to close the cavity and enables the key insertion. To achieve this goal, one must apply a collaring force that exceeds the target coil pre-stress value. Once the keys are inserted and the collaring force removed, the collared coil assembly relaxes resulting in a significant coil pres-tress reduction with respect to the peak. The collaring technique was successfully applied to all industrial productions of Nb–Ti accelerator magnets up-to date, as once the upfront investment on the tooling has been made, it is easy to control and leads to reproducible results. As mentioned earlier, a relevant example of application of the collaring technique can be found in the case of the Nb-Ti LHC main dipole magnets. For these magnets, during the collaring process, the coils were compressed up to 115 MPa - 135 MPa, depending on the method used to apply the load in the press (the maximum allowed peak stress was set to avoid the damage of the conductor insulation). Upon key insertion and force removal, an average compression of about 70 MPa for the inner layer and 75 MPa for the outer layer were obtained. After magnet cool-down to cryogenic temperature, the pre-stress in the coil decreased by between 25 MPa to 40 MPa as a result of the higher thermal contraction coefficient of the coils with respect to the outer structure [32].

When dealing with new high-field Nb$_3$Sn magnets, where larger electro-magnetic forces are present and where controlling peak stress during assembly becomes critical [24], the B&K technique does offer some advantages; as for instance, the possibility to have a gain of coil pre-compression at cryogenic temperature. Nevertheless, the technique has never been yet implemented in an industrial environment.

At the moment of writing this manuscript, the B&K technique has been applied to the assembly of single aperture Nb$_3$Sn dipole and quadrupole magnet models and demonstrators. A non-exhaustive list of the main projects that rely on the technique follows next:

- For superconducting dipole magnet models and demonstrators: SM, LR, SD, RD, HD, SMC, RMC, FRESCA2, E-RMC and RMM [33]-[45].

- Regarding quadrupole magnet models and demonstrators: SQ, TQS, LQ, HQ and MQXF [46]-[50].

In addition, "inflatable bladders" have been as well commonly used in assemblies of large structures, as for instance, in the cold mass preparation of the ATLAS barrel toroid magnets [51].

*1.6. Content of this Manuscript*

In this manuscript we will present a comprehensive experimental and numerical modelling campaign, aimed at optimizing the use of pressurized bladders for high field magnets, such as the HL-LHC MQXFB quadrupole magnets at CERN. Results on short model magnets and long prototypes have shown that the bladder inflation step may lead to a certain coil stress overshoot (reaching a maximum of ≈ 40 MPa in full-length MQXFB quadrupole magnet prototypes, which corresponds to 50 % of the measurable warm pre-load target), similarly to the case of the collaring technique described above. The overshoot comes from the fact that in order to insert the interference keys, the opening of an additional geometrical clearance is required. Such an extra displacement is needed to deal with geometrical imperfections, manufacturing tolerances and to reduce the frictional resistance during key insertion. This requirement becomes more important for long magnets, where geometrical deviations due to assembly tolerances can be more severe than for short models.

With our work, we prove that the B&K assembly procedure can be optimized to eliminate the coil over-stress during bladder operations thanks to the proper positioning of the bladders within the structure. The removal of the coil stress overshoot allows to meet the requirements on peak stress limits with contingency. Furthermore, the conclusions and lessons learnt presented here can be applied to any other high field magnet design based on the bladders and keys assembly process. Note that the stress overshoot is something that cannot be straightforwardly avoided in the collaring technique (unless relying on the so-called "tapered key" technique described in [30]).

In Section 2 of the paper, a brief introduction to the MQXF magnet design is provided. Section 3 explain the use of pressurized bladders for the assembly of MQXF magnets at CERN. Finally, the optimization study and the discussion of the results are presented in Sections 4 and 5 respectively.

## 2. MQXF, the Nb$_3$Sn low-$\beta$ quadrupole for HL-LHC

The High Luminosity Large Hadron Collider (HL-LHC) project at CERN aims at upgrading the existing LHC machine to increase its luminosity by a factor of five beyond the current nominal parameter, and its integrated value by a factor of ten in about a dozen years after commissioning [5].

An important part of this upgrade is focused on the insertion regions of the LHC, in particular near Point 1 (ATLAS experiment) and Point 5 (CMS experiment) [5], which calls for an upgrade of the so-called inner triplets. Among various components to be upgraded, the installation of new low − $\beta$ quadrupole magnets is envisaged. The new magnets, called MQXF [6], [52]-[58], will replace the today's present Nb-Ti quadrupole magnets that were provided as in-kind contribution to the LHC by USA and Japan [59]-[60].

The MQXF low − $\beta$ quadrupole magnets for HL-LHC are designed to produce a gradient of 132.2 T m$^{-1}$ in a 150 mm single aperture, and will operate in a superfluid helium bath at 1.9 K with a conductor peak field of 11.3 T in nominal conditions (see Table 2, which shows as well the expected large e.m. forces acting in the coil windings).

The MQXF magnets come in two lengths (magnetic lengths): 4.2 m for the MQXFA magnets and 7.2 m for the MQXFB magnets. MQXFA magnets, which are an in-kind contribution from the US-HL-LHC Accelerator Upgrade Project (AUP), and MQXFB magnets, which are manufactured at CERN, share the same design and manufacturing procedures. In the case of AUP, two MQXFA magnets are assembled into a common cold-mass and cryostat for Q1/Q3 [61]-[62], whilst at CERN, the longer MQXFB version is assembled in a single configuration for Q2a/Q2b [29]. Except for features related to the outer stainless steel shell welding and fixed point implementation, the AUP and CERN cold masses and cryostats share also similar designs. The first magnet design was proposed in the mid-1990s [63]-[64], while the initial magnet development was carried out as a joint effort between CERN and the previous US Accelerator Research Program (LARP) [29], [65]-[67].

Table 2: Main magnet parameters.

| Parameter | Unit | |
|---|---|---|
| Coil clear aperture diameter | mm | 150 |
| Magnet (LHe vessel) outer diam. | mm | 630 |
| No. turns in layer 1 / 2 (octant) | | 22 / 28 |
| Operational temperature, $T_{op}$ | K | 1.9 |
| Magnetic length (Q1-Q3) / (Q2) | m | 4.2 / 7.2 |
| Operational gradient, $G_{nom}$ | T/m | 132.2 |
| Operational current, $I_{nom}$ | kA | 16.23 |
| Operational conductor peak field | T | 11.3 |
| $I_{op}$ / $I_{ss}$ at 1.9 K | % | 76 |
| Stored energy at $I_{nom}$ | MJ/m | 1.15 |
| $F_x$ / $F_y$ (per octant) at $I_{nom}$ | MN/m | +2.41 / -3.38 |
| $F_z$ (whole magnet) at $I_{nom}$ | MN | 1.15 |
| $F_\vartheta$ layer 1 / layer 2 (per octant) | MN/m | -1.58 / -2.14 |

As already anticipated, the mechanical design is based on the aluminum shrinking cylinder concept and relies on the bladders and key technology for coil pre-loading. The four two-layer Nb$_3$Sn coils produced by the wind, react and impregnate technique are surrounded by Aluminum collars (6061 aluminum, manufactured by CNC milling). The function of the collars is that of a spacer, as they do not provide any mechanical pre-load to the coils. A G11 pole key is then placed on top of the pole. Pads are bolted around the collared coil assembly, defining the so-called coil-pack. Specifically, laminated low-carbon steel pads are used for the magnet straight section (thick laminations produced by CNC milling are placed every 200 mm for MQXFB, with the rest of the length occupied by thin laminations produced by fine blanking). In the magnet ends, thick pad laminations (CNC milled) are made of 304CO steel to decrease the peak field.

The coil-pack sub-assembly is connected to the surrounding low-carbon steel yoke by the low-carbon steel masters, which provide the necessary slots for the keys and the bladders. The yoke is as well a laminated structure made of thin (produced by fine blanking) and thick (produced by combining water jet cutting, CNC milling and EDM wire cutting) laminations. Both pad and yoke assemblies seek for a packing factor of 100 %.

Finally, segmented aluminum cylinders (7075 aluminum, precisely manufactured using open die forging, CNC turning and CNC milling) surround the rest of structural components. Welded stainless steel outer shells and end domes complete the cold mass and serve as LHe containment.

In the longitudinal direction, the axial pre-load is provided by a system of thick end-plates (Nitronic 40) and tie rods (stainless steel 304). The four tie rods (one per quadrant) are inserted into the yoke and pre-tensioned at room temperature to provide a total longitudinal force of ≈ 1.1 MN at 1.9 K (on four quadrants and per magnet side), corresponding to 92 % of the expected axial component of the Lorentz load [68].

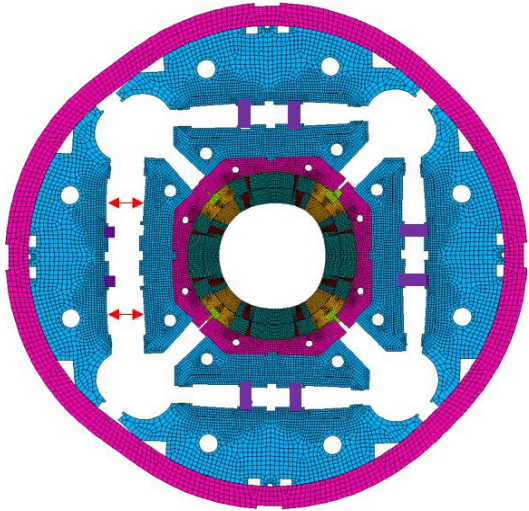

Figure 2: Simulated deformed geometry during a "quadrant by quadrant" bladder operation. The overshoot happens in the coils located opposite to the active bladder. Red arrows show the local pressure application.

In order to monitor the magnet mechanical behavior during the assembly and operation phases, MQXF model and prototype magnets at CERN have been equipped with mechanical instrumentation. Namely, electrical strain gauges and optic fibers (FBG) have been used to measure the structural deformation of the system at selected locations [69]-[72]. The magnet cross section is illustrated in Figure 1; the position of the instrumentation is shown with round markers.

## 3. The assembly of MQXF magnets

### 3.1. MQXFS experience

The collaboration between CERN and LARP/AUP on MQXF started with an initial development phase, after the design activities, during which a number of so-called short model magnets have been produced to qualify, and whenever required, optimize the design and fabrication process. MQXF short models, referred to as MQXFS, feature a magnetic length of 1.2 m and share the same cross-section as that of MQXFA and MQXFB.

The way in which the MQXFA and MQXFB magnets are assembled has been established based on the short model experience [73]. In this regard, the review of the magnet assembly procedure results of paramount importance for our later development and it is explained next.

### 3.2. Main Assembly Steps

#### 3.2.1. Coil size measurements

Systematic dimensional measurements are always performed at the end of the MQXF coil manufacturing process. For MQXFB, they are done by placing the coil on simple supports on a marble and using a FARO® arm portable coordinate measurement machine, CMM. The goal of these measurements is to obtain an evaluation of the coil geometry. Parameters as the azimuthal and radial coil size (given as deviations with respect to the reference coil design) are extracted [74].

#### 3.2.2. Coil Shimming

The coil size information is used to design the so-called "shimming plan" for each magnet, which compensates for the geometrical variability among coils. The current MQXF coil assembly baseline relies on the use of customized azimuthal shims at the coil mid-planes, used to bring all coils to the average size of the largest one. No compensation is provided for azimuthal coil size variations along the coil axis. In the radial direction, shims are used (at the inner surface of the collars) to correct the shift in coil outer radius imposed by the final coil size after mid-plane shimming. Once the azimuthal and radial shims are established, the final coil-pack can be assembled accordingly.

This shimming strategy differs from the one used at the time of the LHC Nb–Ti coils, which employed dimensional measurements carried out on a so-called Young's modulus press. The approach has also been applied to $Nb_3Sn$ impregnated coils [75], but pure CMM measurements have been deemed to be safer. Given the highly non-linear and hysteretic behavior of the coil strain-stress curve [76]-[79], the employed procedure may be subjected to further refinement in future $Nb_3Sn$ programs.

Finally, and since it will become important for next sections of the manuscript, we need to highlight here an interesting signature of the MQXFB coils produced so-far. For the mentioned coils, the measured azimuthal size has been found to be regularly larger in their central section (a phenomenon referred to as "coil belly" [29]). The difference with respect to the coil ends is typically in the order of 150 μm (total arc-length). Maintaining the baseline strategy of employing mid-plane shims of constant thickness, such a longitudinal variation in coil azimuthal size could still be compensated by using loading keys of variable size. The use of these keys has been studied for MQXFB and it is explained in details in [28]. However, it has been decided not to implement this feature in the assembly of prototype and series magnets.

#### 3.2.3. Coil-pack centering

In order to perform the magnet pre-load, the coil-pack is firstly introduced in the previously assembled yoke-cylinder structure. Before proceeding with the final loading, the coil-pack needs to be properly positioned in the center of the structure, in good contact with the iron yoke. This operation is usually referred to as "centering" and consists in the installation of low-interference keys in all quadrants. In the case of MQXF, the design key size that engages all parts in contact (without any interference) is 13 mm. For centering, 13.2 mm keys are inserted in all quadrants.

#### 3.2.4. Magnet pre-load

The final pre-load of the magnet takes place after the coil-pack has been centered. The interference in all quadrants is steadily increased from 13.2 mm up to the target value, which for MQXF has been assigned based on the test results of short model magnets. On this subject, different pre-load levels were explored during the production of MQXFS model magnets. The target pre-load for the long magnets was chosen as that of the short models showing the best performance during cold test. Instead of an interference value (which can differ due to the actual geometry of the coils and magnet structure), the objective pre-load is better defined in terms of the Al-cylinder and coil strain (or stress) measured by the mechanical instrumentation at the selected locations. The explored parameter space in terms of average pole azimuthal stress (over the 4 coils) ranged from −30 MPa in MQXFS6 up to −100 MPa in MQXFS5 [80]. The final target values selected for MQXFA [81] correspond to MQXFS4 and are set at (−80 ± 8) MPa of average azimuthal compression in the coil winding poles and (58 ± 6) MPa of average azimuthal tension in the Al-cylinder (4 angular measurements at each longitudinal position). As it will be shown next, these values will be modified for MQXFB magnet after our optimization study.

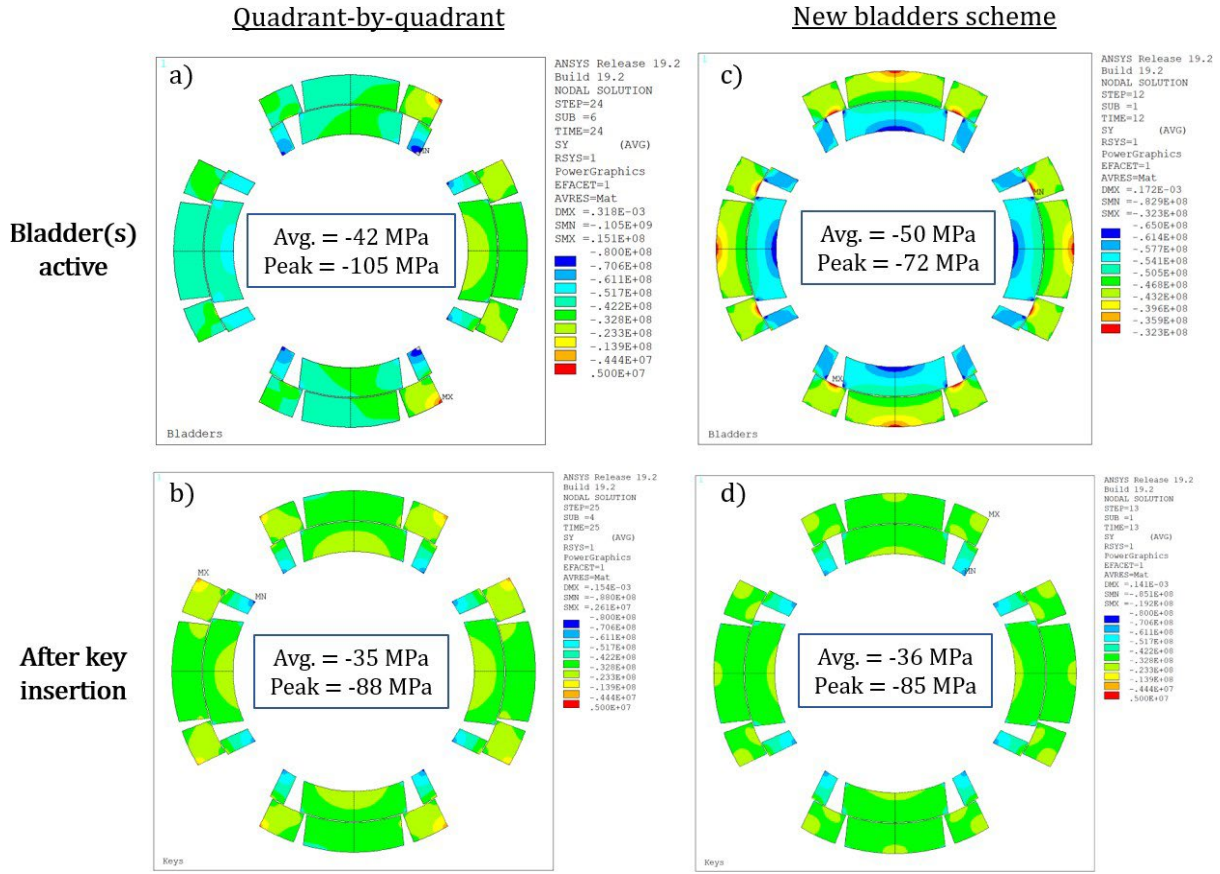

Figure 3: Computed azimuthal stress distribution in the superconducting coils for the quadrant-by-quadrant case (left column) and for the new symmetric procedure (right column). The first row corresponds to the moment when the bladders are inflated, before the insertion of the last loading key. In the case of the quadrant-by-quadrant model, the active bladder acts on the left of the picture. The second row shows the results after key insertion. The peak and average stresses are reported in the center of each image. The units for the stress values in the legend is [Pa]. Simulations have been performed for a clearance, $c$, of 300 µm. The FE model confirms that the experimental measurements performed in the winding pole are representative of the stress value for the inner layer pole turn conductor.

For the last two stages, the centering and actual pre-load, two main alternatives have been historically explored in terms of the bladder - key insertion operations: (i) the full bladder strategy, and (ii) the quadrant-by-quadrant option.

In (i) all bladders in the structure are pressurized at the same time, *i.e.* the bladders from each quadrant are all connected in parallel to the same pressure. Such water supply scheme creates an equal and symmetric force distribution within the magnet, which preserves the symmetry of the system. Conversely, in (ii), the bladder operation is done by pressurizing each quadrant one by one. In this scheme, the force distribution within the structure is asymmetric as illustrated in Figure 2.

The major advantage of the quadrant-by-quadrant scheme is that the pressure required to further enlarge the clearance between masters (when gradually increasing the key size) can be reduced [28]. Consequently, the risk of failure due to the burst of the bladders is significantly decreased. This strategy (quadrant-by-quadrant) has been used in the assembly of all MQXFS model magnets at CERN, both for the centering and magnet pre-load. For MQXFA magnets, a mixed procedure combining the full bladder and quadrant-by-quadrant approaches is currently used: at the moment of writing this paper, the full bladder strategy is employed up to a key size of ≈13.15 mm. From thereon, the quadrant-by-quadrant procedure is used. For MQXFB magnets, the three first prototypes followed the short model experience, using exclusively the quadrant-by-quadrant method. As explained thereafter, a new procedure has been devised and will be applied from now on, also retrofitted to the last short model magnets assembled.

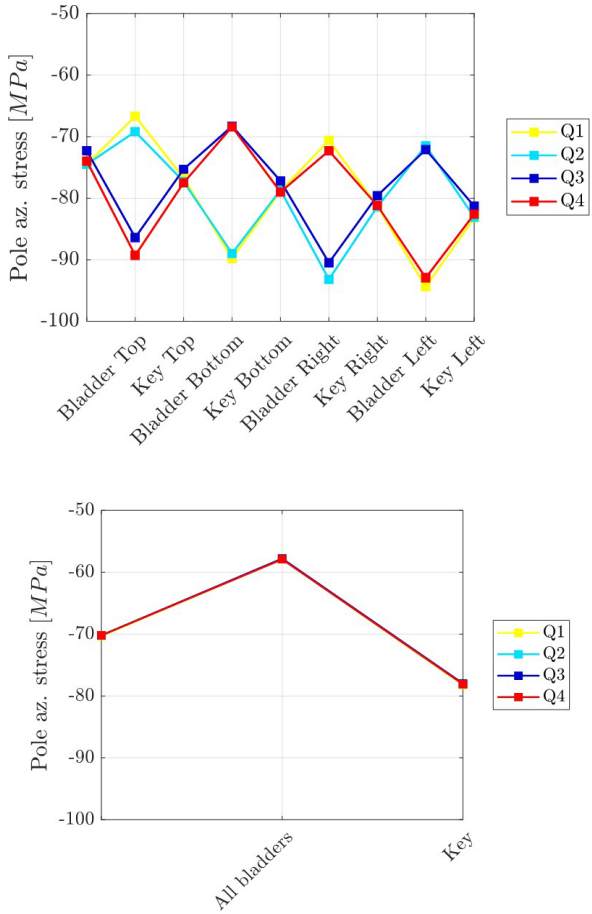

Figure 4: Stress value predicted by the model at the position of the mechanical instrumentation (coil winding poles) for the insertion of the last loading keys. The top plot shows the case of the quadrant-by-quadrant operation, where four bladder-key operations are needed. The bottom one shows the results for the symmetric procedure using the cooling hole bladders. Q1, Q2, Q3 and Q4 refer to the four magnet quadrants.

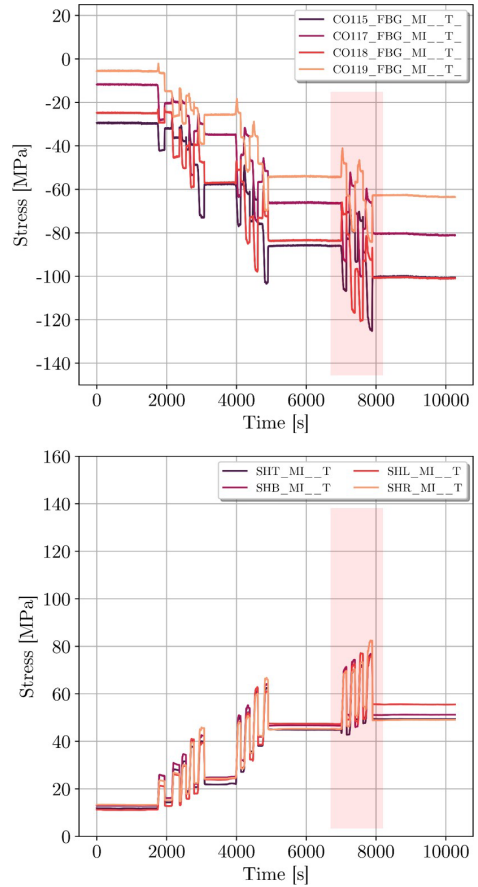

Figure 5: Winding pole (top) and Al-cylinder (bottom) azimuthal stress values (inferred from mechanical strain experimental measurements) during the assembly of the magnet prototype MQXFBP3. The measurement position corresponds to the magnet center. The red rectangles highlight the last four bladder - key operations, performed to install the 13.7 mm keys in all quadrants.

### 3.3. The "Overshoot Concern" during MQXF Assembly.
#### 3.3.1. Origin of the Overshoot

An important consequence of using interference keys in the B&K technique is that a certain clearance needs to be created by the bladder force in order to insert the keys. As previously introduced, this additional opening is needed in order to deal with geometrical imperfections, assembly tolerances and to reduce the frictional resistance during key insertion. As an example, to insert a key of a given size, $k_s$, the distance in between masters shall be $k_s + c$, where $c$ is the required clearance.

To understand the importance of this additional opening in the 7.2 m-long MQXFB assembly, let us focus on the simplest case where the full bladder method is used and no bending or non-linear effects are present. In this case, the need for reaching a clearance between masters that is larger than the final size of the interference key, translates into a larger pre-compression force seen by the coils during the bladder operation. Indeed, in this case, the displacement created by the bladders can be seen as an equivalent key of size $k_s + c$. When the bladder pressure is released, the coil pre-load relaxes from its peak to its final value. This phenomenon is similar to what happens in the case of a collared-coil assembly upon key insertion completion and release of the collaring press force.

The case of the quadrant-by-quadrant operation is more complex and requires a detailed simulation of the full process using Finite Element modelling techniques. In such a case, the non-symmetric loading of the structure introduces important bending and non-linear frictional effects. In addition, before having introduced the last interference key in all quadrants, a mixed state composed of different key thickness in the four quadrants appears. The simulation of the MQXF quadrant-by-quadrant assembly process using ANSYS APDL® is shown in Figure 3 - images (a), (b) and in Figure 4 (top). The plots show the computed azimuthal stress distribution map in the superconducting coils (Figure 3 - a, b) and the corresponding calculated local stress values at the position of the mechanical instrumentation (Figure 4 - top). In the simulation, the geometry of the coils and structural parts corresponds to the reference design one, without geometrical imperfections; the interference has been adjusted to obtain −80 MPa at the end of the magnet pre-load, and the target clearance for key insertion is 300 µm. The model predicts that when the bladders in one quadrant are inflated, the two poles opposite to the active bladders experience an increase of their azimuthal compression. The predicted overshoot in pole stress is approximately 11 MPa. On the contrary, the az. compression decreases in the two poles placed next to the active bladders. In terms of the computed azimuthal peak stress in the coil blocks (Figure 3 - a, b), the overshoot equals 17 MPa. Therefore, the peak coil load during magnet assembly happens during the inflation of the last bladders due to the combination of a non-symmetric structure deformation (bending and non-linear effects) and the clearance requirement.

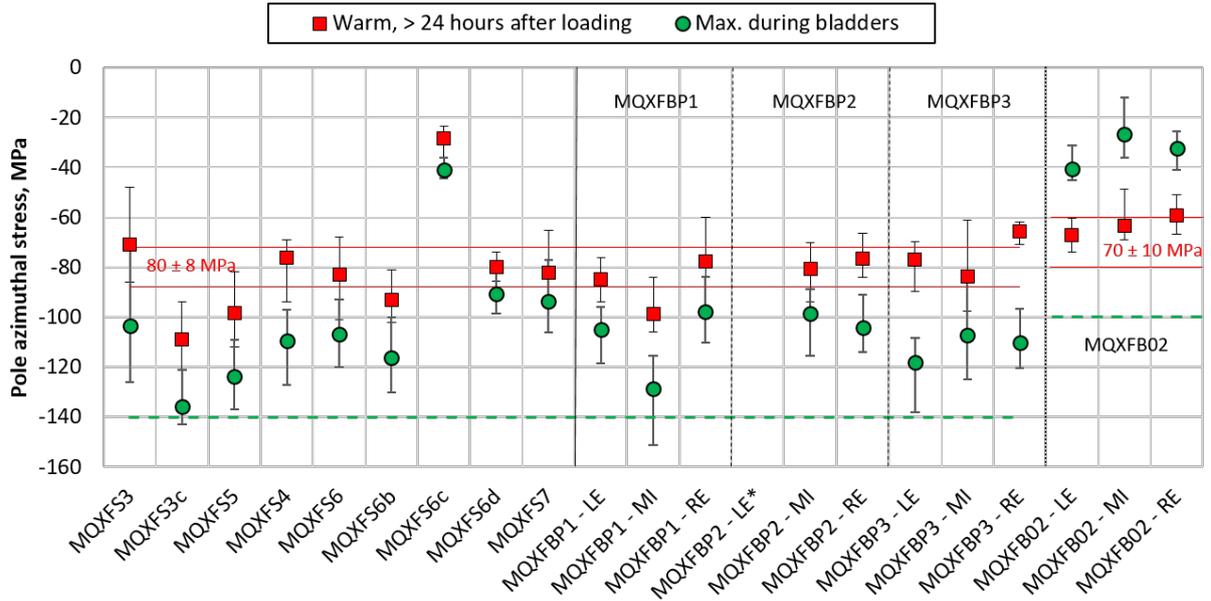

Figure 6: Pole azimuthal stress results for the most relevant MQXF magnets assembled at CERN. Red square markers depict the average azimuthal stress in the winding pole for each magnet and measurement position. Green dots show the average peak stress value reached during the last bladder operation. The corresponding maximum and minimum absolute values are provided with the bars. Note that for MQXFS7, the values reported correspond to the initial magnet assembly using the quadrant-by-quadrant bladder operations. In MQXFBP2, the results on the LE side of the magnet have been removed from the plot due to a large coil imbalance after centering (under investigation). Red lines define the winding pole stress targets. The green ones establish the peak stress requirements. LE stands for Lead End, MI for Middle and RE for Return End.

### 3.3.2. The 7.2 m-long MQXFBP3 Example

Figure 5 shows one example from the experimental mechanical measurement performed during the assembly of the third MQXFB quadrupole prototype magnet (MQXFBP3). The plots depict the derived azimuthal stress (from the actual strain measurements, assuming a bi-axial stress state) on each winding pole (top) and on the Al-cylinder (bottom). Red rectangles point to the last four bladder-key operations. The measurement location corresponds to the middle of the magnet (MI). With MQXFBP3 assembled using the quadrant-by-quadrant strategy, a pole stress overshoot in the order of 20 MPa is seen for all signals. The final stress average can be computed from the plot as ≈ −80 MPa. Expanding the analysis to the three MQXFBP3 magnet positions where mechanical instrumentation is available (recall Figure 1), the peak azimuthal stress on the coils during bladdering is −138 MPa in the Lead End side (LE), −125 MPa in the Middle (MI) and −120 MPa in the Return End side (RE). In previous prototypes, the maximum az. stress measured on a coil during bladdering were −151 MPa for MQXFBP1 and −136 MPa for MQXFBP2 (notice that this last value is not shown in Figure 6 due to a large coil imbalance after centering that prevents the calculation of average values).

### 3.3.3. Assembly Data Summary

Figure 6 summarizes all the relevant coil stress data for the MQXFS and MQXFB magnets assembled so far at CERN. Red square markers show the average azimuthal stress in the winding pole after assembly for each magnet and measurement position. Green dots represent the average peak stress value reached during the bladder operations. The corresponding maximum and minimum absolute values are provided with bars. For long MQXFB magnets, the local azimuthal stress in the four winding poles may reach a peak of −140 MPa or above during bladder operations. As shown by the graph, when the bladder pressure is released, the average azimuthal stress obtained on MQXFB magnets is instead on target at approximately −80 MPa. The mean pole stress overshoot spans from 20 MPa to 30 MPa in the middle of the three first prototypes. In the Return End (RE) and Lead End (LE) measurement positions, the overstress reaches up to 45 MPa.

The pole overshoot is also present for short model magnets. If the first three prototype magnets (assembled using virgin coils) are compared to the short models produced in the same way (from MQXFS3 to MQXFS6), it can be seen that the average over-stress in the winding pole is similar for both long and short magnets. This is especially true when comparing the MQXFS short magnet models with the center of long prototypes, where the four coils composing the magnet are at their largest azimuthal size [29]. The average overshoot for MQXFS magnets (MQXFS3 to MQXFS6) ranges from 24 MPa to 33 MPa. Short magnet models have reached stable performance at nominal current, and beyond, with peak pole azimuthal stresses observed during magnet assembly of ≈ −140 MPa (MQFS5). Cold test results have shown that this level of stress does not prevent from reaching 90 % of short sample limit, and above, in many cases.

To conclude this section, additional interesting information for the B&K technology can be extracted from Figure 6. Neglecting geometry imperfections in the magnet structure, it follows from the MQXF design that longitudinal variations in coil size must translate into variations of the final winding pole stress (after key insertion, due to the modification of the local interference). The underlying assumption for this behavior is that both the thickness of the loading keys and shimming layers remain constant along the length. As a result, for MQXFB, one would expect the azimuthal compression in the winding pole to be systematically larger in the center of the magnet (biggest coil size, see Section 3.1). The available results in Figure 6 for MQXFBP1, MQXFBP2 and MQXFBP3 are well in line with the expected behavior for the final measurements after magnet pre-loading (red squares). During the bladder operations, instead, bladders are expected to act as "smart shims" that inflate under the exerted internal pressure while compensating for potential geometrical imperfections. In this respect, a similar stress level in the winding pole should be obtained for the three longitudinal stations in MQXFB prototypes during bladder inflation. This expected coil behavior is not seen in the results of Figure 6. As mentioned in the introduction, it should not be forgotten that the mechanical measurements performed in the winding pole (based on FBG sensors for full-length prototypes) correspond to a localized strain field. Bending effects, for instance, may appear in the system due to the actual coil - collar contact condition, affecting the azimuthal stress value at the measurement position. The origin of this inconsistency is currently under investigation. A detailed sensitivity analysis on the impact of coil geometry variations in the MQXF stress state, including the cases of the three first prototypes, can be found in [28].

## 4. Optimization Study: The Use of Additional Bladders in the Yoke Cooling Hole Channels to Release Coil Peak Stress

### 4.1. Conceptual Procedure and Analysis

Based on the assembly experience and on the cold test results from the first MQXFB prototypes [29], a detailed investigation campaign on the MQXFB magnet production was launched at CERN in Spring 2021. The goal of this campaign was to carry out a thorough root cause analysis of the performance limitation of the first two MQXFBP1&2 prototypes and derive possible risk mitigation measures. The analysis identified three possible root causes related to (1) cold mass assembly, (2) magnet assembly and (3) coil manufacturing. As a part of (2), it was decided to review the pre-load targets and to study as to minimize the peak stress level seen by the superconducting coils during assembly at room temperature. In order to reach this objective, the removal of the stress overshoot during bladder operations was strongly pursued. It should be noted that according to the magnet finite element computations, the coil peak stress does not occur neither during assembly at room temperature nor during energization, but at the end of cool-down in the inner layer coil pole block. This effect could be alleviated by introducing a notch in the winding pole. However, it was deemed more appropriate to avoid this notch so as: (i) to limit the displacement of the pole turn conductor corner during magnet energization (which increases in presence of the notch, as a result of the reduction in the pole rigidity), (ii) to avoid a discontinuity in the pole tip geometry towards the coil extremities (a smooth transition to a solid pole should be engineered at the coil ends), and (iii) to allow for the installation of the mechanical instrumentation used for pole strain assessment [70].

The cure to the stress overshoot was found thanks to the analysis of the force distribution in the system when bladders are pressurized and by imitation to the process followed at CERN for the loading of other magnets and demonstrators [82]. Following the original MQXF bladder positioning selected at the beginning of the project, the bladder operations exert a simultaneous load in the coils and Al-cylinders. This force distribution eventually results in the appearance of the undesired stress overshoot. Without any modification to the magnet design, if the bladder force were to be applied exclusively at the interface between iron yoke parts, one could obtain a net gain in cylinder tension without any part of this force being applied to the coils. In simpler words, the solution would consist in placing additional bladders in the iron yoke cooling holes that would open up the cylinder without exerting any force onto the coil-pack. As just stated, such a solution is inspired by the previous experience coming from the assembly of racetrack model magnets (SMC, RMC, E-RMC and RMM [39]-[43]) and the FRESCA2 dipole magnet [44]-[45] at CERN. For these magnets, additional bladders located in the iron yoke are used in combination with the coil-pack ones. By doing so, one can reach the required assembly force while keeping the bladder pressure at reasonable values, i.e. < 400 bar (see next paragraphs).

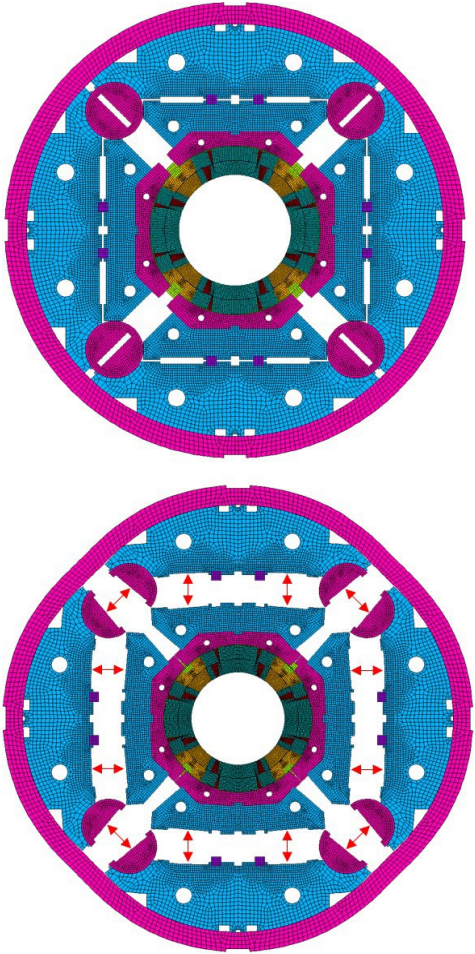

Figure 7: Proposed new bladder system (top). Simulated deformed geometry during a symmetric bladder operation with new bladders in the magnet cooling holes (bottom). Red arrows show the active bladders.

The practical implementation of this modification in MQXF was investigated using the two-dimensional FE model of the magnet. The desired force distribution in the system could be achieved by introducing the additional bladders as shown in Figure 7. These auxiliary bladders act exclusively on the yoke and Al-shrinking cylinders, unloading the coil-pack.

For the operation of the bladders, it was decided to switch to the full-bladder pressurization strategy (all quadrants active at the same time, including the cooling hole bladders). The reasons behind this choice are two-fold: (i) to reduce the complexity and number of bladder cycles per magnet assembly by a factor of four, and (ii) upon completion of the centering phase, to keep a symmetric operation of all bladders so as to allow the correct coil-pack position when the new cooling hole bladders are active. The main drawback of the full operation could be, however, the need of excessive bladder pressures. Nevertheless, the performed FE simulations predict that the latter can be kept below the acceptable 400 bar level, when targeting for the reference magnet pre-load. Such a pressure value has been set based on practical experience, and it can be seen as an acceptable level that avoids the most-common failure mode of the bladders.

For the centering phase, the quadrant-by-quadrant strategy was kept (no use of new cooling hole bladders), as it has been deemed more appropriate. In this case, the main advantage comes from its asymmetric nature, which can help to correct initial deviations in the coil-pack positioning with low bladder pressures. Note that for quadrant operations, at the beginning of centering (when no interference is present), the only reaction force opposing the coil-pack displacement comes from the weight of the coil-pack and the frictional interaction with the structure.

The right side of Figure 3, with images (c) and (d), shows the azimuthal stress distribution in the superconducting coils during the last bladder operation (c) and after the insertion of the interference keys (d) for the revised procedure. The figure compares these new results to the original quadrant-by-quadrant option, shown on the left side. Contrary to the quadrant-by-quadrant case (with a computed peak azimuthal stress in the coil block, $\sigma_{y\ peak} = -105$ MPa), the use of the new cooling hole bladders results into a decrease of the coil peak stress during bladder operations ($\sigma_{y\ peak} = -72$ MPa). Since the same pressure is used for all bladders, the average compressive azimuthal stress in the coil is instead larger than in the quadrant by quadrant method ($-50$ MPa compared to $-42$ MPa during bladder pressurization). This compressive value could be adjusted by using a larger pressure in the cooling hole bladders than in the coil-pack ones. However, this modification has not been considered necessary, as the average value remains well within the tolerable limits for the conductor.

The expected azimuthal stress at the mechanical instrumentation position is shown in the bottom of Figure 4. During final bladder pressurization the pole stress is actually 20 MPa lower than after the final key insertion according to the FE model.

A very instructive plot used in the assembly of B&K based magnets is the so-called transfer function graph, which plots the azimuthal stress level in the coil winding poles as a function of the azimuthal stress in the external Al-cylinder. It gives the essential information on how much azimuthal force is provided by the cylinder and which force reaches the superconducting coils. Interestingly, according to the 2D FE model, the use of the new bladder procedure translates into a modification of the magnet transfer function (see Figure 8, which plots the computed transfer function for the different assembly methods). For the same Al-cylinder stress, the azimuthal stress level in the winding pole is lower when all bladders are operated at the same time. Detailed investigations of this behavior have revealed that the reduction in coil stress is a result of the global frictional interaction between the different parts of the structure. The latter modifies the local stress seen in the coil winding poles. The simulation also predicts that when the magnet is cooled-down to cryogenic temperature, the friction effects disappear and the stress state is equivalent to that of a magnet assembled with a quadrant-by-quadrant strategy.

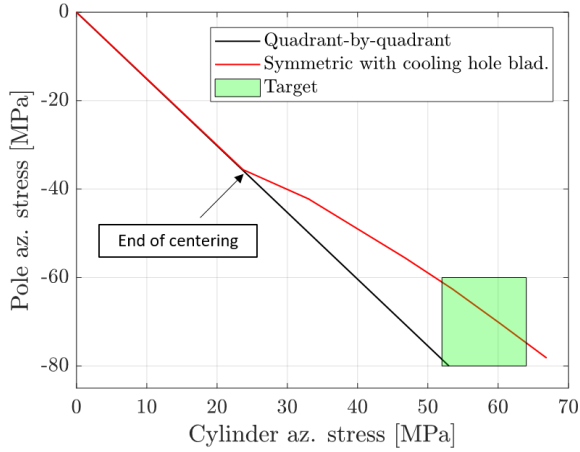

Figure 8: MQXF transfer function for the two assembly alternatives. In both cases, the centering of the magnet is performed using the original procedure. The green rectangle delimits the new targets for MQXF, see Section 4.4. No contact at the pole-key level is assumed [28].

## 4.2. Experimental validation

Prior to the application of the new procedure to the assembly of a full-length MQXFB quadrupole magnet, the method was firstly put in practice in a full-length mechanical assembly test (MQXFBMT3) and in a short quadrupole magnet model (MQXFS7e).

### 4.2.1. The 7.2 m-long MQXFBMT3

In details, the MQXFBMT3 mechanical assembly test was a magnet assembled with four non-conform $Nb_3Sn$ coils, whose objective was to refine the MQXFB magnet assembly procedures and to qualify the implementation of the new bladder scheme. In addition, the magnet was heavily instrumented, including two supplemental longitudinal positions for the mechanical instrumentation. Being a so-called mechanical assembly, the magnet was disassembled few weeks after the end of the magnet loading, no cool-down to cryogenic temperature was performed.

Figure 9 shows the results of the mechanical measurements carried out in the central position of the MQXBFMT3 assembly test. As desired, when the new bladder scheme is employed, the peak winding pole azimuthal stress disappears during bladder pressurization. This behavior is obtained at the expense of a larger tension in the Al-cylinder, as shown in the bottom of Figure 9. The azimuthal stress in the Al-cylinder reaches 130 MPa in MQXFMBT3, in comparison with the 80 MPa seen for MQXFBP3 (quadrant-by-quadrant, Figure 5). Despite this increase, the values are well below the yield limit of the material (420 MPa at 300 K [83]). Note that sharp geometries have been removed from MQXFB Al-cylinders to avoid undesired stress concentration problems.

The absolute peak stress in the coils now appears after the insertion of the last loading keys and not anymore during bladder operation. Focusing on the final assembled state, the experimental results confirm as well that with the new procedure, the stress imbalance between coils remains of the same order of magnitude than for previous magnet loadings (±15 MPa, see also Figure 6).

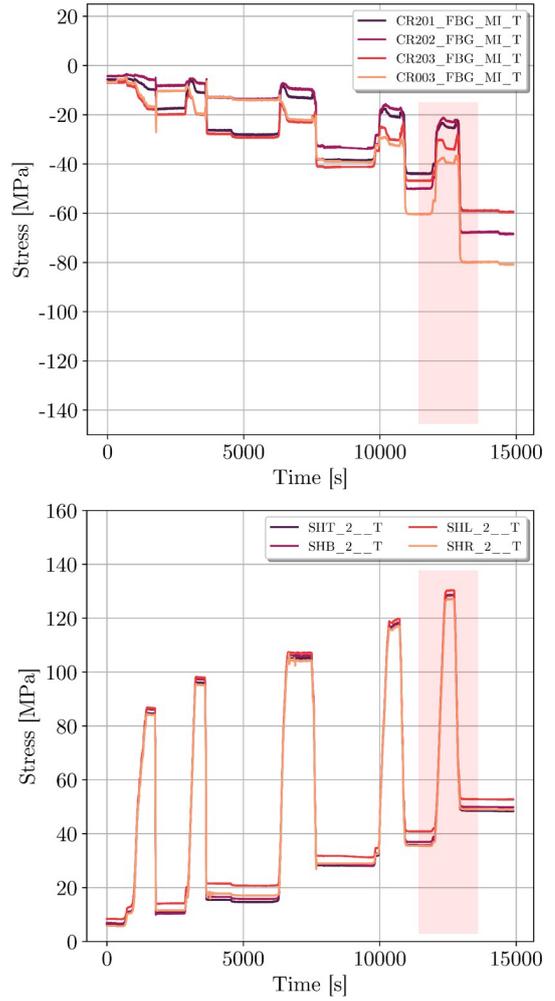

Figure 9: Winding pole (top) and Al-cylinder (bottom) azimuthal stress during the assembly of MQXFBMT3 mechanical test. The red rectangles highlight the last bladder operation.

In global terms, the MQXFBMT3 mechanical test campaign was successfully completed. No showstopper was found for the practical implementation of the new bladder scheme. Despite a larger displacement under pressure (bladder stroke) when compared to their iron master counterparts, the new cooling hole bladders worked reliably and confirmed the feasibility for their use.

## 4.3. MQXFS7e

MQXFS7e short model is the fifth assembly of the MQXFS7 magnet. With the objective of further verifying the Finite Element model predictions, the magnet was fully unloaded from its previous configuration (MQXFS7d) and then re-assembled twice. The first assembly (MQXFS7e1) was performed using the original quadrant-by-quadrant procedure. Upon completion of the magnet pre-load, it was unloaded immediately. In the second assembly (MQXFS7e2), the magnet was pre-loaded using the newly optimized method with additional bladders in the yoke cooling holes. Note that both assemblies (MQXFS7e1 and MQXFS7e2), reproduced the magnet pre-load target of MQXFS7d, previously cold tested. To verify the mechanical behavior at cryogenic temperature of MQXFS7e2, the magnet was cold tested in the same conditions than MQXFS7d.

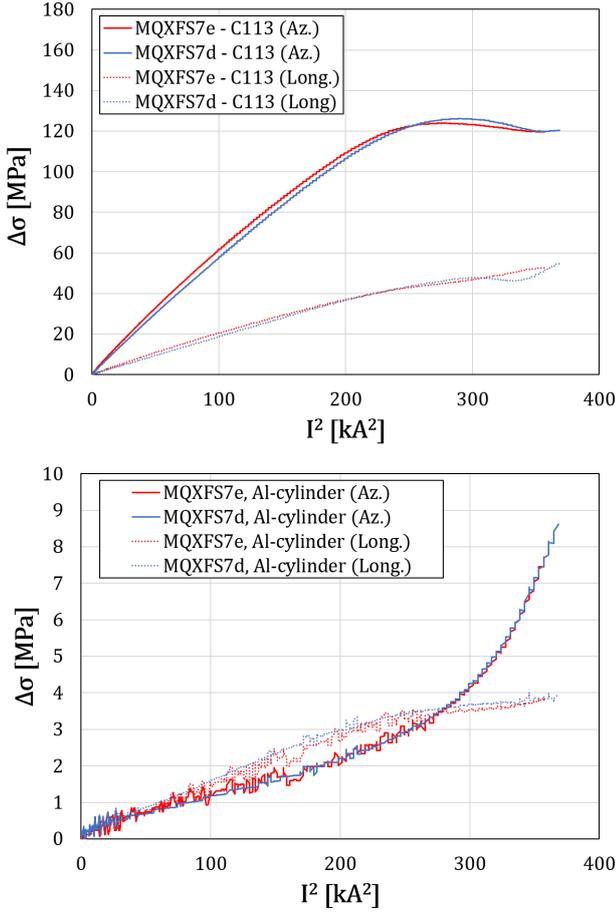

Figure 10: Stress variation comparison during magnet powering at 1.9 K in the winding pole of one of the coils (top) and in the Al-cylinder (bottom). Red color represents the case of MQXFS7e, assembled with the new cooling hole bladders. Blue color is used for MQXFS7d, assembled with the old quadrant-by-quadrant bladder procedure. Solid lines plot the delta azimuthal stress, computed by subtracting the initial stress value at cryogenic temperature (before magnet energization) to the subsequent values obtained during the current ramp. Dotted lines plot the longitudinal equivalent signals for the same coil and Al-cylinder.

The goal of this full campaign was three-fold: (i) to verify that the new loading procedure had no detrimental effect on magnet performance, (ii) to further study the impact of the procedure in the magnet transfer function, and (iii) to confirm that as predicted by the model, the magnet mechanical behavior at cryogenic temperature is independent of the assembly sequence, even if the winding pole azimuthal stress value is different at warm.

The responses to (i) and (iii) were found during the MQXFS7e cold test phase [84]. The magnet, assembled with the new procedure, reached the same current level as its previous version pre-loaded using the quadrant-by-quadrant bladder scheme (MQXFS7d). In addition, the mechanical measurements performed at cryogenic temperature confirmed that the mechanical behavior at cold is identical for both strategies. This last statement is supported by the two plots of Figure 10, which show the stress variation during a current ramp for the winding pole and Al-cylinder. Solid lines are used for the azimuthal stress direction, dotted lines for the longitudinal one. The tests show the same response during cold magnet powering, despite a measurable difference in the transfer function at warm (not shown here and explained in next paragraph).

The answer to (ii) is graphically represented in Figure 11. The top plot shows the comparison on the pole azimuthal stress achieved by both loading strategies, as a function of the interference key size. The bottom graph shows the same information regarding the cylinder azimuthal stress. As predicted by the model, the same cylinder stress results into a lower local stress in the winding pole in the case of the new bladder strategy. The pole stress difference for a cylinder azimuthal tension of 40 MPa (13.7 mm key in this magnet) is 14 MPa. The value is in very good agreement with the expected difference based on the FE model, 15 MPa (see Figure 8). As already mentioned, this difference disappears at cryogenic temperature.

All together, the obtained experimental results confirm the FE model outcome and provide the necessary confidence for the application of the new procedure in the assembly of full-length MQXFB prototypes and series magnets.

### 4.4. Application to the MQXFB02 Magnet Assembly.

Having validated the new pre-loading procedure both in the short magnet model and in the full-length MQXFBMT3 test, the technique was then applied for the assembly of a real full-length prototype magnet.

Thanks to the introduction of the auxiliary bladder system, more stringent pre-load targets and stress limits could be established. The new values seek to minimize as much as possible the risk for conductor degradation due to excessive conductor strain. They also take into consideration the change in the magnet transfer function during loading and the expected final state at cold. The exact values are listed hereinafter:

- The average azimuthal Al-cylinder stress in the magnet (average of all the sensors located in the three longitudinal positions) shall be: +58 ± 6 MPa.

- The average azimuthal stress in the coil winding poles (average of all the poles, including the three longitudinal positions) shall be: −70 ± 10 MPa. This corresponds to a reduction of 10 MPa with respect to the original target. However, the pre-load at cold would be the same.

- The peak azimuthal stress in any of the coil winding poles (at any longitudinal position) must be always within −100 MPa.

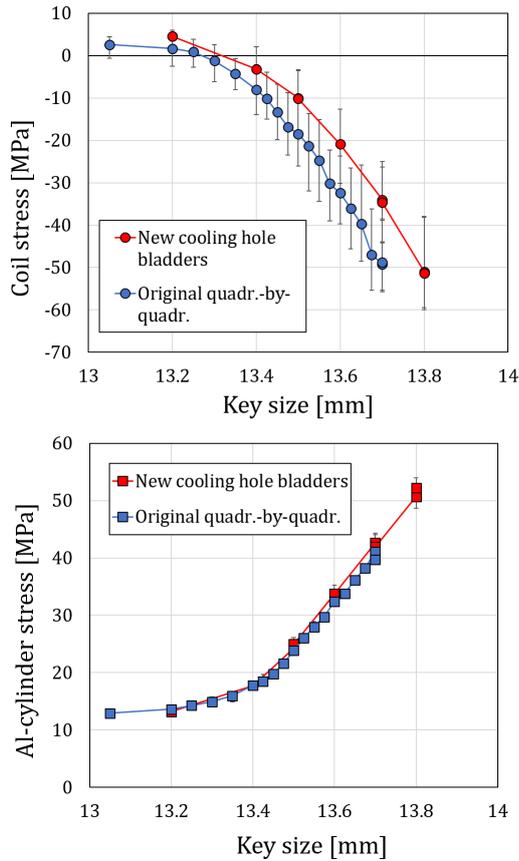

Figure 11: Winding pole (top) and Al-cylinder (bottom) azimuthal stress as a function of the loading key size during the two MQXFS7e magnet assemblies at room temperature. Markers show the average values for all coil / Al-cylinder instrumentation, while lines extend in between maximum and minimum values.

The results from the magnet assembly are shown in the summary plot of Figure 6. All the targets and stress limits have been fulfilled. The magnet was assembled with a maximum absolute peak stress value (winding pole) of −74 MPa. Comparing to previous prototypes, a peak stress reduction in the order of 70 MPa has been achieved.

## 5. Discussion

The main advantage of the new procedure lies on the elimination of the coil stress overshoot during bladder operations. Nevertheless, other important lessons learnt from our study must be mentioned as well.

First, it has been explained in previous sections, that the peak coil azimuthal stress at ambient temperature is now found after the insertion of the interference keys, and corresponds to the final assembled state of the coils. Such a stress value can be extrapolated based on the coil stress achieved with previous key sizes. See Figure 11, where after a first non-linear part during which the cylinder tension is still kept by the yoke keys (used during the yoke-cylinder assembly to keep both parts together), the increment in azimuthal stress becomes approximately linear. By carefully monitoring the stress increase as a function of the key size, a reliable estimation of the peak stress for the next key size can be obtained. Although the prediction capability is not unique for this method (the AUP collaboration adapts the quadrant-by-quadrant sequence based on the previous bladder peak stress attained), it can be seen as a powerful tool for guaranteeing that the thresholds mentioned in the last section are achieved and offers a reliable quality control procedure.

Second, an equally important development for the B&K application to MQXFB magnets has been the improvements in the bladder technology. Initially, the bladders were produced by welding two stainless steel sheets over the full bladder perimeter. When subjected to large displacements, this type of bladders were prone to burst due to the failure of the welded connections. Improved bladders are today produced by rolling stainless steel tubes. In this case, only two welding operations (at each bladder extremity) need to be performed. The reliability of the new bladders, which have been used for all prototype magnets except for MQXFBP1, is greatly increased thanks to this new production process. As an example, during the assembly of MQXFBMT3, MQXFBP2, MQXFBP3 and MQXFB02 magnets, none of the bladders failed or had to be replaced (with a maximum pressure attained of 400 bar).

To close the discussion section, it has been shown that the need for a certain clearance and the geometrical imperfections within the system play an important role in the B&K assembly. When the bladders are pressurized, the gap created for the key insertion needs to be always such that it overcomes the worst point (in terms of geometry variability) along the full magnet. This is especially important for long magnet units. With a dramatic reduction of the peak load values in the coils, the optimized method for MQXFB has enabled the successful assembly of long magnets with more stringent peak stress thresholds. This has been achieved while preserving the same pre-load target at cryogenic temperature.

## 6. Conclusions

The assembly of $Nb_3Sn$ accelerator magnets requires the careful control of the mechanical loads experienced by the superconducting coils, in order to decrease the risk of conductor degradation. This manuscript presents the results of a full experimental and numerical modelling campaign aimed at optimizing the use of water pressurized bladders for magnet assembly. The results of this campaign, whose main outcomes are enumerated thereafter, will be applied to the assembly of the 7.2-m-long MQXFB quadrupole series magnets manufactured at CERN:

1. During the assembly of both MQXFS magnet models and long MQXFBP1-3 prototype magnets at CERN, a systematic stress overshoot in the winding pole was observed during bladder operations. The average overshoot was of the order of 20 MPa for both magnet versions. When monitoring the maximum local winding pole peak azimuthal stresses, short and long MQXF quadrupoles have reached values in excess of −140 MPa during magnet assembly (for a target average azimuthal stress of −80 MPa).

2. With the objective of eliminating the coil over-stress, an optimized bladder strategy has been proposed. The solution, which has become the baseline for the assembly of MQXFB magnets at CERN, consists in the temporary installation of additional bladders in the iron yoke cooling holes. These new bladders enable to increase the Al-cylinder tension while reducing the coil peak compression.

3. Finite element simulations show that this new bladder scheme removes completely the stress overshoot in the system. Furthermore, with the use of the cooling hole bladders, the peak mechanical load in the superconducting coils happens after the insertion of the loading keys. The predicted peak azimuthal stress in the coil blocks during bladder operation is now −72 MPa (for the MQXF target pre-load), which is a reduction of 33 MPa with respect to the expected value for the original bladder scheme (−105 MPa).

4. The model predictions have been confirmed by the practical implementation of the new bladder strategy in both a short model magnet and in a full-length mechanical assembly test. Based on the positive results, the optimized bladder system was finally applied to a real full-length MQXFB prototype. The absolute experimental peak stress in the coil winding poles for this magnet remained as low as $\sigma_\vartheta = -74$ MPa.

5. The new bladder solution has enabled the definition of more stringent stress limits for the MQXFB magnets without a detrimental impact on magnet performance (demonstrated in the short MQXFS7e magnet model).

6. The principle on which this optimization is based can be applied to other magnet designs relying on the B&K assembly process and offer interesting perspectives for future applications.

## 7. Acknowledgements

The authors would like to acknowledge the full technical staff of the TE-MSC group involved in the construction and cold powering test of MQXFB magnets. Special thanks to E. Takala for the initial development of the full MQXF 2D Finite Element model and for pointing out the large peak stress values observed on the first MQXFB prototypes. The authors are also indebted for numerous exchanges and constructive confrontations with their AUP counterparts.

## References

[1] H. Brechna, "High field superconducting magnet systems", Springer-Verlag Berlin Heidelberg (1973).

[2] K.H. Mess, S. Wolff, and P. Schmuser, "Superconducting Accelerator Magnets.", World Scientific (1996).

[3] M. N. Wilson, "Superconducting Magnets", Oxford University Press, Oxford (1983).

[4] A. Devred, "Superconducting magnets for particle accelerators and storage rings," Wiley Encyclopedia of Electrical and Electronics Engineering, Vol. 20, pp. 743-762 (1999)

[5] L. Rossi and O. Bruning, "The High Luminosity Large Hadron Collider," Advanced Series on Directions in High Energy Physics: Volume 24. World Scientifc (2015).

[6] E. Todesco et al 2021 Supercond. Sci. Technol. 34 053001.

[7] A. Devred, "Practical low temperature superconductors for electromagnets," CERN-2004-006 (Yellow Report), CERN, Geneva, Switzerland, 12 July 2004

[8] J. W. Ekin, "Unified scaling law for flux pinning in practical superconductors: I. Separability postulate, raw scaling data and parametrization at moderate strains", Supercond. Sci. Technol. 23, 083001 (2010).

[9] J.W. Ekin, "Strain scaling law for flux pinning in practical superconductors. Part 1: Basic relationship and application to Nb$_3$Sn conductors," Cryogenics, volume 20, issue 11, November 1980, Pages 611-624 (1980)

[10] J. W. Ekin, Experimental Techniques for Low-Temperature Measurements: Cryostat Design, Material Properties, and Superconductor Criticalcurrent Testing. New York: Oxford University Press, 2006.

[11] D. Arbelaez, A. Godeke, and S. O. Prestemon, "An improved model for the strain dependence of the superconducting properties of Nb$_3$Sn," Supercond. Sci. Technol., vol. 22, no. 2, 2009, Art. no. 025005.

[12] B. Bordini, P. Alknes, L. Bottura, L. Rossi, and D. Valentinis, "An exponential scaling law for the strain dependence of the Nb$_3$Sn critical current density," Supercond. Sci. Technol., vol. 26, no. 7, 2013, Art. no. 075014.

[13] L. Bottura et al., "$J_c(B, T, \epsilon)$ Parameterization for the ITER Nb$_3$Sn Production," IEEE Trans. Appl. Supercond., Vol. 19, No. 3, (2009).

[14] C. Calzolaio et al., "Electro-mechanical properties of PIT Nb$_3$Sn wires under transverse stress: Experimental results and FEM analysis," Supercond. Sci. Technol., vol. 28, no. 5, pp. 1–11, 2015.

[15] B. Bordini, P. Alknes, A. Ballarino, L. Bottura, and L. Oberli, "Critical current measurements of high-Jc Nb$_3$Sn rutherford cables under transverse compression," IEEE Trans. Appl. Supercond., vol. 24, no. 3, Jun. 2014, Art. no. 9501005.

[16] J. E. Duvauchelle, B. Bordini, J. Fleiter, and A. Ballarino, "Critical current measurements under transverse pressure of a Nb$_3$Sn Rutherford cable based on 1 mm RRP wires," IEEE Trans. Appl. Supercond., vol. 28, no. 4, Jun. 2018, Art. no. 4802305.

[17] Cheggour, N., Stauffer, T.C., Starch, W. et al., "Implications of the strain irreversibility cliff on the fabrication of particle-accelerator magnets made of restacked-rod-process Nb$_3$Sn wires," Sci Rep 9, 5466 (2019).

[18] P. Ebermann et al., "Influence of transverse stress exerted at room temperature on the superconducting properties of Nb$_3$Sn wires," Supercond. Sci. Technol., vol. 32, p. 095010, 2019, doi: 10.1088/1361-6668/ab2e51.

[19] L. Gamperle, et al., "Determination of the electromechanical limits under transverse stress of high-performance Nb$_3$Sn Rutherford cables from a single-wire experiment," Phys. Rev. Research 2, 013211 – Published 26 February (2020).

[20] J Ferradas Troitino et al 2021 Supercond. Sci. Technol. 34 035008

[21] P. Ebermann et al., "Influence of transverse stress exerted at room temperature on the superconducting properties of Nb$_3$Sn wires," Supercond. Sci. Technol., vol. 32, p. 095010, 2019, doi: 10.1088/1361-6668/ab2e51.

[22] P. Ebermann et al., "Irreversible degradation of Nb$_3$Sn Rutherford cables due to transverse compressive stress at room temperature", Supercond. Sci. Technol., vol. 31, (2018).

[23] H. Felice et al., "Performance of a Nb$_3$Sn Quadrupole Under High Stress," in IEEE Transactions on Applied Superconductivity, vol. 21, no. 3, pp. 1849-1853, June 2011, doi: 10.1109/TASC.2010.2090116.

[24] P. Ferracin et al, "Mechanical Analysis of the Collaring Process of the 11 T Dipole Magnet," IEEE Transactions on Applied Superconductivity Volume: 29, Issue: 5, August 2019.
[25] G. Lenoir, private communication, 2021.
[26] K. Puthran et al., "Onset of mechanical degradation due to transverse compressive stress in Nb3Sn Rutherford cables," Submitted for publication.
[27] S. Caspi et al., "The use of pressurized bladders for stress control of superconducting magnets," IEEE Transactions on applied superconductivity, Vol. 11, No. 1, March 2011.
[28] E. Takala et al., "On the mechanics of MQXFB- the low-beta quadrupole for the HL-LHC," 2021 Supercond. Sci. Technol. 34 095002.
[29] S. Izquierdo et al., "Status of the MQXFB Nb3Sn quadrupoles for the HL-LHC," Submitted for publication.
[30] A. Devred, T. Bush, et al., "About the mechanics of SSC Dipole magnet prototypes," AIP Conference Proceedings, No. 249(2), pp. 1309-1374 (1992; Invited.)
[31] P. Fessia, D. Perini et al., "Selection of the Cross-Section Design for the LHC Main Dipole," IEEE Trans Appl Supercond Vol 10 No 1 pp. 65-68, 2000
[32] Bruning, Oliver; Collier, Paul; Lebrun, P ; Myers, Stephen ; Ostojic, Ranko; Poole, John; Proudlock, Paul; "LHC Design Report" CERN-2004-003-V-1 (Yellow Report), CERN, Geneva, Switzerland, 2004
[33] R. Hafalia et al., "A new support structure for high field magnets," IEEE Transactions on Applied Superconductivity, vol. 12, issue 1, March 2002.
[34] K. Chow, et al., "Mechanical Design of a High Field Common Coil Magnet," 1999 Particle Accelerator Conference, New York, NY, March, 1999.
[35] S.A. Gourlay, et al., "Design and Fabrication of a 14 T, Nb3Sn Superconducting Racetrack Dipole Magnet," IEEE Transactions on Applied Superconductivity, Vol.10, pp. 294 – 297, 2000.
[36] S. Caspi, M. Fong, S. Gourlay, R. Hafalia, A. Lietzke and J. O'Neill, "RD3 Structure," SC-MAG 725, Lawrence Berkeley National Laboratory, June, 2000.
[37] P. Ferracin et al., "Mechanical Analysis of the Nb3Sn Dipole Magnet HD1," IEEE Transactions on Applied Superconductivity, vol. 15, issue 2, June 2005.
[38] H. Felice et al., "Design and test of a Nb3Sn subscale dipole magnet for training studies," IEEE Trans. Appl. Supercond., vol. 17, pt. 2, pp. 1144–1148, Jun. 2007.
[39] F. Regis et al., "Mechanical Design of the SMC (Short Model Coil) Dipole Magnet," IEEE Trans. Appl. Superconduct., vol. 20, no. 3, Jun. 2010.
[40] J.C. Perez et al., "16 T Nb3Sn Racetrack Model Coil Test Result," IEEE Trans. Appl. Superconduct., vol. 26, no. 4, Jun. 2016.
[41] S. Izquierdo et al., "Design of ERMC and RMM, the base of the Nb3Sn 16 T magnet development at CERN," IEEE Trans. Appl. Supercond., vol. 27, no. 4, Jun. 2017, Art. no. 4002004
[42] J.C. Perez et al., "Construction and Test of the Enhanced Racetrack Model Coil, First CERN R&D Magnet for the FCC" IEEE Transactions on Applied Superconductivity, Volume: 32, Issue: 6, September 2022.
[43] E. Gautheron et al., "RMM," IEEE Trans. Appl. Superconduct., vol. X, no. X, Jun. 2022.
[44] A. Milanese et al., "Design of the EuCARD high field model dipole magnet FRESCA2", IEEE Trans. Appl. Supercond., vol. 22, no. 3, Jun. 2012.
[45] E. Rochepault et al., "Mechanical analysis of the FRESCA2 dipole during assembly, cool-down, and training," IEEE Trans. Appl. Supercond., vol. 28, no. 3, Apr. 2018, Art. no. 4002905.
[46] P. Ferracin et al., "Assembly and tests of SQ02, a racetrack quadrupole magnet for LARP," IEEE Trans. Appl. Supercond., vol. 17, no. 2, June 2007.
[47] S. Caspi et al., "Design and analysis of TQS01, a 90 mm Nb3Sn model quadrupole for the LHC luminosity upgrade based on a key and bladder assembly," IEEE Trans. Appl. Supercond., vol. 16, no. 2, pp. 358–361, Jun. 2006.
[48] G. Ambrosio et al., "LARP long Nb3Sn quadrupole design study," IEEE Trans. Appl. Supercond.,IEEE TRANSACTIONS ON APPLIED SUPERCONDUCTIVITY, VOL. 18, NO. 2, JUNE
[49] G. L. Sabbi et al., "Design studies of Nb3Sn high-gradient quadrupole models for LARP," IEEE Trans. Appl. Supercond., vol. 17, no. 2, pp. 1051–1054, Jun. 2007.
[50] H. Felice et al., "Design of HQ—A High Field Large Bore Nb3Sn Quadrupole Magnet for LARP" IEEE Trans. Appl. Supercond., vol. 19, no. 3, June 2009.
[51] J.M Rey et al., "Cold Mass Integration of the ATLAS Barrel Toroid Magnets at CERN, " IEEE TRANSACTIONS ON APPLIED SUPERCONDUCTIVITY, VOL. 16, NO. 2, JUNE 2006.
[52] E. Todesco et al., "Design studies for the $low-\beta$ quadrupoles for the LHC luminosity upgrade," IEEE Trans. Appl. Supercond., vol. 23, no. 3, Jun. 2013, Art. no. 4002405.
[53] E. Todesco et al., "A first baseline for the magnets in the high luminosity LHC insertion regions," IEEE Trans. Appl. Supercond., vol. 24, no. 3, Jun. 2014, Art. no. 4003305.
[54] P. Ferracin et al., "Magnet design of the 150 mm aperture $low-\beta$ quadrupoles for the high luminosity LHC," IEEE Trans. Appl. Supercond., vol. 24, no. 3, Jun. 2014, Art. no. 4002306.
[55] P. Ferracin et al., "Development of MQXF: The Nb3Sn $low-\beta$ quadrupole for the Hilumi LHC," IEEE Trans. Appl. Supercond., vol. 26, no. 4, Jun. 2016, Art. no. 4000207.
[56] G. Ambrosio, "Nb3Sn high field magnets for the High Luminosity LHC upgrade project," IEEE Trans. Appl. Supercond., vol. 25, no. 3, Jun. 2015.
[57] F. Lackner et al., "Status of the Long MQXFB Nb3Sn Coil Prototype Production for the HiLumi LHC," IEEE Transactions on Applied Superconductivity (Volume: 27, Issue: 4, June 2017)
[58] P. Ferracin et al., "The HL-LHC $low-\beta$ Quadrupole Magnet MQXF: From Short Models to Long Prototypes," IEEE Trans. Appl. Supercond., vol. 29, No 5, Aug 2019. Art. no. 4001309
[59] S. Feher et al., "Production and installation of the LHC low-$\beta$ triplets," IEEE Trans. Appl. Supercond., vol. 16, no. 2, pp. 437–440, Jun. 2006.
[60] A. Yamamoto et al., "Production and measurement of the MQXA series of LHC low-$\beta$ insertion quadrupoles," IEEE Trans. Appl. Supercond., vol. 15, no. 2, pp. 1084–1089, Jun. 2005.
[61] R. Bossert et al., "Design of the Fermilab Pre-Series Cold Mass for the HL-LHC Accelerator Upgrade Project," IEEE Transactions on Applied Superconductivity ( Volume: 32, Issue: 6, September 2022).
[62] G. Vallone et al., "A novel design for improving control on the stainless-steel vessel welding process for superconducting magnets," Submitted for publication.
[63] G. Ambrosio, G. Bellomo and L. Rossi Study of a High Gradient, Large Aperture, Nb3Sn Quadrupole for the Low-b Insertions of the LHC Proc. of Fourth European Particle Accelerator Conference, (EPAC 94), London, June 1994, World Scientific publisher, p. 2268
[64] G. Ambrosio, F. Ametrano, G. Bellomo, F. Broggi, L. Rossi, G. Volpini Preliminary Proposal of a Nb3Sn Quadrupole Model for the low b Insertions of the LHC INFN report, N. INFN/TC-95/25, 13 September 1995
[65] S. A. Gourlay et al., "Magnet R&D for the US LHC accelerator research program," IEEE Trans. Appl. Supercond., vol. 16, no. 2, pp. 324–327, Jun. 2006.
[66] P. Ferracin, "LARP Nb3Sn quadrupole magnets for the LHC luminosity upgrade," in Proc. AIP Conf., 2010, pp. 1291–1300, vol. 1218.
[67] Strait, J B; Bru¨ning, Olivier Sim; Caspi, S; Devred, A; Gourlay, S A; Harrison, M; Lamm, M J; Limon, P; Mokhov, N V; Ostojic, R; Peggs, S; Pilat, F; Rossi, L; Ruggiero, F; Sabbi, G L; Sen, T; Gupta, R; Taylor, T; Zlobin, A V; ten Kate, Herman H J, "Towards a new LHC Interaction Region design for a luminosity upgrade, Particle Accelerator," Conference PAC 2003, Portland, OR, USA, 12-16 May 2003 - pp 42-44 0-7803-7739-9 ©2003 IEEE
[68] G. Vallone et al., "Mechanical Design Analysis of MQXFB, the 7.2-m-Long $Low-\beta$ Quadrupole for the High-Luminosity LHC Upgrade," IEEE TRANSACTIONS ON APPLIED SUPERCONDUCTIVITY, VOL. 28, NO. 3, APRIL 2018.
[69] M. Guinchard et al., "Mechanical Strain Measurements Based on Fiber Bragg Grating Down to Cryogenic Temperature – Precision and Trueness Determination" in 26th International Conference on Optical Fiber Sensors, OSA Technical Digest (Optical Society of America, 2018), paper WF85.
[70] M. Guinchard et al., "Techniques of mechanical measurements for CERN Applications and Environment" CERN, Rep. EDMS no. 1064933, 2010.
[71] L. Bianchi et al., "Strain measurements on MQXFS5 superconducting

magnet," CERN, Rep. EDMS no. 1701525, 2018.

[72] A. Chiuchiolo et al., "Strain Measurements With Fiber Bragg Grating Sensors in the Short Models of the HiLumi LHC Low-Beta Quadrupole Magnet" IEEE Transactions on Applied Superconductivity, vol. 28, issue 4, June 2018.

[73] G. Vallone et al., "Mechanical analysis of the short model magnets for the $Nb_3Sn$ low-β quadrupole MQXF," IEEE Trans. Appl. Supercond., vol. 28, no. 3, Apr. 2018, Art. no. 4003106.

[74] J. Ferradas et al., "Applied Metrology in the Production of Superconducting Model Magnets for Particle Accelerators". IEEE Transactions on Applied Superconductivity, vol. 28, issue 3, April 2018.

[75] J.L. Rudeiros et al, "Characterization of the Mechanical Properties of $Nb_3Sn$ Coils," IEEE TRANSACTIONS ON APPLIED SUPERCONDUCTIVITY, VOL. 29, NO. 5, AUGUST 2019.

[76] C. Fichera et al., "New Methodology to Derive the Mechanical Behavior of Epoxy-Impregnated $Nb_3Sn$ Cables," IEEE Trans. Appl. Supercond., vol. 29, issue 7, 2019.

[77] F. Wolf et al., "Effect of Epoxy Volume Fraction on the Stiffness of $Nb_3Sn$ Rutherford Cable Stacks ," IEEE Transactions on applied superconductivity, Vol. 29, No. 5, Aug. 2019.

[78] C. Scheuerlein et al., "Direct measurement of $Nb_3Sn$ filament loading strain and stress in accelerator magnet coil segments," Superconductor Science and Technology, Volume 32, Number 4.

[79] G. Vallone et al., "A methodology to compute the critical current limit in $Nb_3Sn$ magnets," Supercond. Sci. Technol., Accepted for publication (2020).

[80] S. Izquierdo et al., "Performance of a MQXF $Nb_3Sn$ Quadrupole Under Different Stress Level". IEEE Transactions on Applied Superconductivity, vol. 32, issue 6, September 2022.

[81] P. Ferracin et al., "Assembly and Pre-Loading Specifications for the Series Production of the $Nb_3Sn$ MQXFA Quadrupole Magnets for the HL-LHC" IEEE Transactions on Applied Superconductivity, vol. 32, issue 6, September 2022.

[82] N. Bourcey, J.C. Perez, private communication, 2021.

[83] E. Andersen, S. Prestemon et al. "Structural design criteria," US-HILUMI-DOC-909

[84] F.J. Mangiarotti et al., "MQXFS7 test report," EDMS 2748245. Available online: https://edms.cern.ch/document/2748245/1.